%% file: galfxv2.tex
\documentclass[12pt]{article}
        \input defs \begin{document}

        \title {Host Galaxy Effects in the Susy Model for  Supernovae Ia}
%
        \author{ L.
    Clavelli\footnote{Louis.Clavelli@Tufts.edu,\,lclavell@ua.edu}\\ Dept.
    of Physics and Astronomy, Tufts University, Medford MA 02155\\ Dept.
    of Physics and Astronomy, Univ. of Alabama, Tuscaloosa AL 35487}
        \date{\today} \maketitle

        \begin{abstract}

        For more than forty years virtually all work on the theory of
    type Ia Supernovae (SN Ia) has assumed that these explosions were due
    to the transfer of mass to a degenerate star from a partner in a
    binary system. In these binary models, when the mass of one partner
    closely approaches the Chandrasekhar maximum for a stable degenerate
    system, fusion can be initiated and the star explodes. However, a 
    number of long-standing nagging problems and the inability of any
    specific binary model to fit any significant fraction of SN Ia events
    suggest that fusion could instead be triggered by a phase transition 
    in a sub-Chandrasekhar white dwarf star. It is possible that
    remarkable host galaxy effects not considered in
    previous work on phase transition models could point to a specific source 
    of the supernova trigger.  Performing a least $\chi^2$ fit to the delay 
    time distribution to fix parameters, we give predictions from the susy
    phase transition model for the host galaxy effects. In addition we
    discuss a susy insight into the Phillips relation which is basic to
    the cosmological importance of the type Ia supernovae.
    \end{abstract}

        \section[1]{Introduction}

    An exhaustive review of the status of binary models for type Ia
    supernovae has been recently presented in
    ref.\,\cite{Maoz-Mannucci-Nelemans}. An essential feature of these models
    is that accumulation of matter from a partner star leads to an SN Ia
    explosion near the Chandrasekhar mass, $M_C$. A very recent review of 
    all the binary models from the point of view of progenitor identification
    has also been compiled \cite{Livio}. A conclusion of this latter review
    is that no binary model describes a large fraction of the observed SN Ia.
    In addition, a point made there is that, even if a particular progenitor
    is identified in the future, the question remains as to what happens to
    the other suggested progenitors as the Chandrasekhar mass is approached.  
    Furthermore, if type Ia supernovae have a variety of initial states, one
    must explain how a sufficient degree of homogeneity is produced for 
    cosmological applications.
    
    In contradistinction to the binary models we have noted 
    \cite{SixIndications} that there are at least six major 
    indications that SN Ia may be due not to mass growth 
    to the Chandrasekhar maximum but, instead, to a phase transition in a 
    possibly isolated white dwarf. 
  
        Some years ago, Sim et al. \cite{Sim} showed that, if a suitable
   trigger could be found, the detonation of an isolated white dwarf
   with mass between $0.97$ and $1.15$ could lead to good agreement with
   observed SN Ia properties.  The model of \cite{Biermann-Clavelli},
   \cite{breakdown} in which matter at extremely
    high density undergoes a tunneling phase transition to a background of
    exact supersymmetry (susy) provides such a trigger.  Recently 
    \cite{Blondin} has shown that sub-Chandrasekhar explosions with an 
    undetermined ignition mechanism might also explain the low luminosity
    supernovae.  
           
     The susy model assumes that the exactly
    supersymmetric universe is the ground state of the multiverse and that
    the transition rate to this ground state is enhanced at high density.  In
    such a susy background, initial state fermion pairs would convert to
    boson pairs which, since they are unaffected by the Pauli Principle, 
    would drop to the ground state emitting sufficient energy to trigger
    fusion in the surrounding matter.  Although this model may seem overly
    radical, if further decades pass without any consensus on a precise
    binary progenitor, astronomers may want to tolerate some attention to 
    such phase transition models.    
      
    There have long been observations relating supernova and host galaxy 
    properties. As indicated in \cite{Maoz-Mannucci-Nelemans} the 
    relations are often of low statistical significance 
    and, in some cases, there are even claims differing in the sense of 
    the correlations.  In such a situation it is useful to explore the
    predictions of the susy model with a view to support or disfavor the
    model once definitive data become available.  
    In the previous studies of the susy model cited above, there was no
    consideration of the supernova-host-galaxy correlations which are taken
    up in the present article.  In the binary models, the host-galaxy effects
    could be related to effects on the rate of mass growth although
    chemical composition effects in binary models have also been 
    proposed \cite{Calder}.  Some of the predictions of this latter study
    are, however, not in full agreement with data.  For instance, 
    the binary model predicts a rapidly decreasing nickel mass 
    as a function of metallicity while the data shows constant
    or increasing mass in Ni$^{56}$. If, on the other hand,
    an isolated dwarf can explode as in the phase transition models, the 
    host-galaxy effects must be due to the composition of the dwarf and its 
    relation to galactic age and metallicity.

    In the susy model, as laid out in \cite{Biermann-Clavelli},
    \cite{breakdown}, and \cite{SixIndications}, every white dwarf has a
    natural mass-dependent lifetime.  In a restricted range of mass, the
    lifetime ranges from a small fraction of a gigayear to gigayear scales.
    Below a certain mass, the lifetime of the white dwarf is longer than the 
    current age of 
    the universe so it is effectively stable.  In this model, the 
    delay time distribution (DTD)
    from birth of the white dwarf when fusion ceases to its supernova
    explosion and the ejected mass distribution (EJMD) are foldings of the
    white dwarf mass distribution with a phase transition probability.  
    Each of these could be metallicity dependent as discussed
    below but only the latter is an intrinsic property of the susy model.  
    
    In the binary models, the DTD and EJMD are multi-parameter functions of 
    the binary mass distribution and orbital parameters.  At present there
    are few observational constraints on these parameters so predictions
    are dependent on model dependent simulations.  A prime
    parameter is the fraction of white dwarfs in binary systems.  In the double
    degenerate (DD) scenario this fraction must, most likely 
    \cite{Maoz-Mannucci-Nelemans}, 
    be above $50\%$ although no high mass binary white dwarf 
    systems have as yet been observed.  If this test is passed, if the other
    counter-indications \cite{SixIndications} to binary models are overcome,  
    and if the fits to the DTD and EJMD are equal or better than those of the 
    phase transition model, one might prefer the binary model as 
    possibly being simpler. 

    In the following section 
    we discuss the composition of the white dwarfs and their
    mass distribution.

    \section[2]{The White Dwarf Mass Distribution}

    In the phase transition model the rates are proportional to the
    single white dwarf production rate for which good data are available.  
    White dwarf binaries, in principle, are counted twice in this 
    distribution. 
    
    \begin{figure} [th] \centering \includegraphics[scale=0.85]{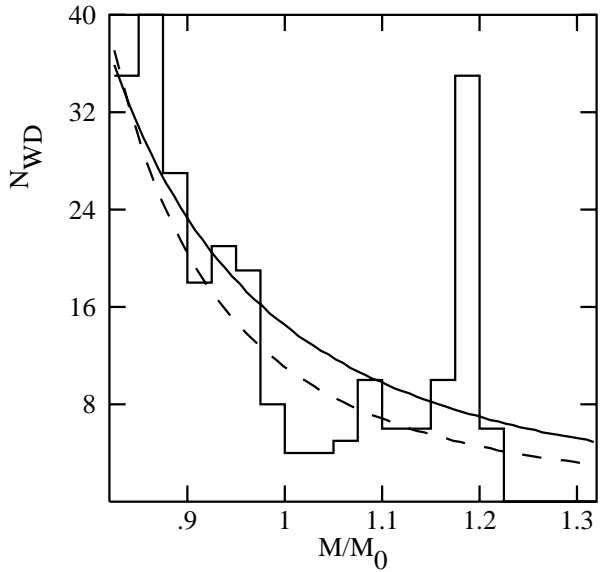}
    \caption{White Dwarf production rate for average to high metallicity
    environments (solid curve) and low metallicity environments (dashed
    curve) compared to white dwarf masses as observed in the low
    metallicity parts of the Milky Way. The solid curve is an 
    expanded view of the high
    mass tail of the full Salpeter type fit to the white dwarf mass
    distribution shown in ref.\,\cite{breakdown} .} 
    \label{IMF5C}
    \end{figure}

    The Salpeter initial mass function with a linear relation between the
    main sequence mass and the resulting white dwarf mass gives an excellent
    fit to the observed hot white dwarf mass distribution between the peak at
    $0.6\, M_\odot$ and $0.8\,M_\odot$ as shown, for example, in 
    \cite{breakdown}. The hot white dwarf mass distribution
    is taken to approximate the birth mass distribution.  However, as can
    be seen in fig.\,\ref{IMF5C}, the Salpeter initial mass function 
    overpredicts the white dwarf mass distribution for masses above $0.8$.   
    The white dwarf mass is a monotonically increasing function of the mass
    of the parent main-sequence star \cite{Williams}.  We assume, as seems
    reasonable, that the higher mass parent stars are of higher temperature
    and hence burn to elements of higher atomic number.  These higher $Z$
    elements are then passed on to the white dwarf as well as to the 
    ambient interstellar medium.  Higher mass white dwarfs are, therefore,
    associated with higher metallicity environments and are produced less
    efficiently in low metallicity environments leading to the deficit 
    relative to the solid curve seen in fig.\,\ref{IMF5C}.  
    Main sequence stars in the range from 8 to 11 solar masses produce
    white dwarfs \cite{Nomoto}, \cite{Woosley} that are primarily 
    oxygen-neon-magnesium mixtures
    as reviewed, for example, in \cite{cronodon}.  Main sequence
    stars below 8 solar masses produce mainly carbon-oxygen dwarfs
    with masses below $0.9 M_\odot$ 
    which form the great majority of observed white dwarfs.  
    White dwarfs near the Chandrasekhar mass may also  
    have an iron core \cite{Panei}, \cite{Madej03}, \cite{Provencal}.
    
    The ``metallicity" in a stellar neighborhood is a variously defined
    measure of the amount of higher atomic number elements in the system.  A
    commonly used single parameter measure, normalized to zero for the sun,
    is 
    \be 
        [Fe/H] = \log_{10} (f_{26}/f_{1}) - \log_{10} (f_{26}/f_{1})_\odot \quad .
    \label{IronContent} 
    \ee 
    Here, $f_{Z}$ is the fraction of atoms having atomic number $Z$.
    The solar abundances are such that the second term has the value $0.02$.
    $[Fe/H]$ is a convenient proxy for a more physically relevant 
    expression which must depend on 
    other heavy elements in addition to iron.  Since iron is inert to 
    fusion, the expression eq.\,\ref{IronContent} nevertheless highlights the 
    expectation that, if fusion
    is the sole energy source, the nickel production and supernova energy 
    release decrease  
    with heavy element abundance contrary to empirical indications (See,
    for example, figure 2 of \cite{Calder}). 
   
      In the thin disk of the Milky Way, which includes the sun, $-0.5 <[Fe/H]
    < 0.3$ while in the thick disk $[Fe/H]$ is typically lower 
    \cite{Carroll-Ostlie}.  In the central bulge of the Milky Way 
    and in galaxy clusters higher
    metallicities are plentiful due, presumably, to a rich history of
    supernovae which efficiently convert low mass elements to high mass
    elements and seed the environment with elements of high atomic 
    number. Mathematically, $[Fe/H]$ ranges over the entire real axis
    as the fraction of iron or hydrogen goes to zero while the atomic
    number averages are positive and limited. 
    
    High concentrations of iron are typically produced together with high
    concentrations of other high Z elements.  [Fe/H] thus 
    monotonically increases with
    the logarithm of the average atomic number of 
    elements in a galactic neighborhood.
    \be
         \overline{Z} = \sum Z_i f_i \quad ,
    \ee 
     In a toy galaxy containing only hydrogen and a small amount of iron, 
    $[Fe/H]$ is related to the average atomic number $\overline{Z}$ 
    in the environment by
    \be
         [Fe/H] = log \frac{f_{26}}{1-f_{26}} + 0.02 \approx log 
         \frac{\overline {Z}-1}{25} + 0.02 \quad .
    \ee  
       
  The average heavy element atomic number of a main sequence star has, 
  initially, a term equal to that number in the environment.
  The heavier elements should condense at the center and be preferentially
  passed on to the daughter white dwarf.
  Thus, while the heavy element fraction in a stellar neighborhood 
  would be expected to vary linearly as a function of 
  the galactic age at stellar birth, the mean atomic numbers in the 
  white dwarf 
  should vary more than linearly due to the central location of the 
  dwarf at birth. 
  These processes could lead to the correlations noted above 
  \cite{Nomoto}, \cite{Woosley},\cite{Panei}, \cite{Madej03}.

    In summary, white dwarfs with masses above $1.0$ 
    will have an average constituent atomic number above that of oxygen and 
    are produced preferentially in high metallicity
    environments whereas most of the observed white dwarfs
    have a C-O composition and come from low metallicity environments.   
    
    We normalize the white dwarf production rate assuming a constant star
    production rate for $12.8\,\mathrm{Gyr}$.  There is, of course, some
    variation in this number and the rate would be greater in
    star-burst galaxies. The Salpeter initial mass function for a 
    main sequence star falls off as the power $2.35$.  If there is 
    a linear dependence of the white dwarf mass, $M(0)$, as a function 
    of the 
    parent main sequence stellar mass. we would expect the white dwarf
    production rate to vary as
     \be
    F(M(0)) =  a_0 ((M(0) - b_1)/c_1)^{-2.35} \qquad M(0)>0.62 M_\odot
    \label{WDIMF1} 
    \ee 
    with 
    \be a_0 = 5.48\, \frac{12.8 Gyr}{t_g} \mathrm{WD/yr/10^{10}WD} \quad . 
    \label{a_0} 
    \ee
    
    For $0.6\,< M(0)\,<1.35$, the fit shown in 
    ref.\,\cite{breakdown}, as appropriate to the
    birth mass distribution in some average metallicity environment
    and, correspondingly, in some average galactic age $t_g$ took
    \be
    b_1 = 0.478 \qquad c_1 = 0.0903 
    \ee 
        
    In the present fit we take a nominal age of $t_g =12.8$ Gyr to fit
    the data in the delay time distribution (DTD).  Good fits can also be
    found with ages ten to twenty percent lower.  With future high statistics
    data it should be possible to study the DTD as a function of the host
    galaxy age and metallicity.

    In lower metallicity environments  
    the rough fit of fig.\,\ref{IMF5C}
    shown in the dashed curve uses
    \be
     b_1 = 0.59 \qquad c_1 = 0.064
    \label{Newb_1}
    \ee
    which comes from the
    crude fit shown in the dashed curve of fig.\,\ref{IMF5C} to the 
    tail of the observed distribution
    of local white dwarfs .  The spike at $M(0) = 1.2$ is known to be an
    artifact of the data treatment at high masses and should be ignored
    \cite{Madej04} or spread among higher masses.  Other authors discussing
    the variation of the initial mass function are refs.\,\cite{Geha},
    \cite{Kroupa}, and \cite{Meng}.
    The white dwarf sample in the histogram of fig.\,\ref{IMF5C} comes 
    from the local
    low metallicity environment whereas the Salpeter fit shown in the 
    solid line derives primarily from high mass main sequence stars 
    and their supernovae which are clustered at about redshift $0.5$
    and which preferentially produce the high mass dwarfs.   
    Since we are primarily interested in the sense of the metallicity
    dependence and only secondarily in its magnitude,
    we do not attempt to define an explicit Z dependence for $b_1$
    and $c_1$.  Roughly speaking,
    in advance of more precisely discriminating data,
    we consider negative $[Fe/H]$ as low multiplicity and $[Fe/H]>0.5$
    as high metallicity.

\section[3]{Phase Transition Probability in the Susy Model}

    If the ground state of the universe is
    exactly supersymmetric and the current state is one of broken
    supersymmetry, the probability per unit space time volume of vacuum decay
    to the ground state in interstellar space is given by the
    Coleman-DeLuccia \cite{Coleman-DeLuccia} formula: 
    \be \frac{d^2 P}{dtd^3x} = A e^{-B} 
    \ee 
     where the action takes the form 
     \be 
     B = \frac{27\pi^2 S^4}{2 \hbar c \overline\epsilon^3}\quad . 
     \ee 
     $S$ is the surface
    tension of a critical bubble of true vacuum nucleated in the ambient
    false vacuum universe and $\overline\epsilon$ is the average energy
    density difference between the two vacua.  Thus, given a knowledge of
    $\overline{\epsilon}$, the transition probability depends on the two
    parameters $S$ with units of energy per unit area and $A$ with units of
    an inverse space time volume.  In the phase transition theory bubbles of
    all radii are constantly being nucleated in the false vacuum but most of
    them will be immediatey quenched in a competition between the volume
    energy, $4 \pi r^3 \overline\epsilon/3$, tending to expand the bubble and
    the surface energy, $4 \pi r^2 S$, tending to collapse it. In a
    homogeneous vacuum, where the energy difference is equal to its average,
    bubbles with initial radius greater than a critical value 
    \be 
    R_c =\frac{3 S}{\overline\epsilon} 
    \ee 
    will grow at the speed of light to
    complete the vacuum transition.  Presumably, the bubble growth speed and
    the speed of light in dense matter are also equal.
     
    \section[4]{Density Enhancement of the Susy Phase Transition}

    Although it is proven only in lower dimensions
    \cite{Gorsky}\cite{inducedvacdecay}, it is reasonable to assume that,
    in a dense medium,  the phase transition is accelerated. This can be
    naturally implemented by assuming that, in the above formulae, the
    vacuum energy difference is replaced by the total energy difference.
    Thus, in dense media,  the lifetime $\tau(M)$ for a white dwarf of mass
    $M$ is given by 
    \be 
    \frac{dP}{dt} = A\,V(M) \equiv \frac{1}{\tau(M)}
    \label{PTprob} 
    \ee 
    where 
    \be 
    V(M) = \int d^3x \, \, e^{-B} 
    \ee 
    and the action now takes the form 
    \be 
    B = \frac{13.5 \pi^2 S^4}{\hbar c (\overline\epsilon+
    \overline{\Delta \rho(r)})^3} \quad . 
    \ee 
    The critical
    radius or minimum radius of a successful bubble nucleation is 
    \be 
    R_c =\frac{3 S}{\overline\epsilon+\overline{\Delta \rho}}\quad .
    \label{CritRad} 
    \ee 
    Without loss of generality we can factor out of $A$
    the inverse of the maximum of $V(M)$ over all white dwarf masses leaving
    a free parameter with dimensions of inverse time. 
    \be 
    \frac{dP}{dt} =\frac{1}{\tau_0} \frac{V(M)}{V_\mathrm{max}}\equiv 
    \frac{1}{\tau (M)}
    \label{dPdt} 
    \ee 
    The free parameter $\tau_0$ becomes then the minimum
    lifetime of the dwarfs in the sample.
    If $\Delta \rho$ is simply and universally proportional to the
    density as assumed in earlier work we can define a critical density
    $\rho_c$ such that, in dense matter where $\overline{\epsilon}$ is
    negligible, 
    \be 
    B = (\frac{\rho_c}{\rho})^3 \quad . 
    \label{action} 
    \ee 
    In white dwarf physics the natural scale of energy is the solar mass,
    $M_\odot$, and the natural scale of distance is the Earth radius, $R_E$.
    In ref.\,\cite{breakdown}, a good fit to the delay time distribution was
    found with $\rho_c = 7.42\cdot 10^7 \mathrm{g/cm^3} = 9.69
    M_\odot/{R_E}^3$. With this parameter value, the surface tension is such
    that the critical radius in empty space is of the order of galactic size
    making our universe safe for billions of additional years.
    In an inhomogeneous medium, the growth of a critically sized bubble
    will be halted if the critical radius of eq.\,\ref{CritRad} becomes
    greater than the current bubble radius. In an uncompactified superstring
    theory or one compactified on tori the ground state vacuum energy density
    vanishes so that the energy density difference is equal to the vacuum
    energy density in the broken susy phase which is currently measured to be
    \be \epsilon = 3.66\, GeV/m^3 \quad . \ee

      \section[5]{Metallicity Dependence of the Susy Phase Transition}

        In the susy phase transition model the energy density difference
    in dense matter is the energy trapped in high energy levels due to
    the Pauli Principle.  This energy is released in a transition to
    exact susy since Fermion pairs can convert to pairs of degenerate
    Bosons.   The conversion takes place in nuclei on strong interaction
    time scales ($\approx 10^{-24}\, s$) via gluino exchange between
    quarks.  On the electromagnetic time scale ($\approx 10^{-6}\,s$)
    electron pairs can convert into selectron pairs via photino exchange.
      The former conversion has a greater energy release but does not
    immediately affect the electron degeneracy pressure which supports the
    star.  
   
    The energy release per unit mass from electron to selectron
    conversion is equal to the electron internal energy that would be
    released as the star explodes.   The electron to selectron conversion
    in a susy core, therefore, provides no additional energy release
    relative to the standard model once the explosion is triggered. 
    Thus, the energy provided in the susy phase transition model can be
    considered as coming
    entirely from the collapse of the Pauli towers in nuclei.
    The first estimates based on the Fermi gas model indicated $\Delta
    \rho \approx 0.02\,\rho$ some twenty times greater than the fusion energy
    released in carbon.  In the Fermi gas model, the Pauli energy is
    primarily a function only of the atomic number to atomic weight ratio,
    $Z/A$.  Consequently the energy release would be the same for the all the
    low-lying stable nuclei carbon, oxygen, neon, and magnesium. This leads
    to a universal form for the action, $B$. However, the Fermi gas model
    fails to predict the magic numbers in nuclei which are naturally
    accommodated in the nuclear shell model. As we will discuss in this
    article,  the nuclear shell model in combination with the phase
    transition model does predict atomic number dependence and allows for the
    correct sense of the SN Ia-host-galaxy correlation.

    The primary host galaxy properties of interest in supernova physics
    are:\\ 
    1. metallicity\\ 
    2. galaxy age\\ 
    3. galaxy mass\\ 
    4. star formation rate
        
    The itemized galactic properties are strongly
    interrelated.  It can be argued that high mass galaxies have a 
    greater number of high mass stars which will produce a greater
    number of high mass white dwarfs.
    Since elements of high atomic number are produced in
    supernovae,  older galaxies with a longer history of supernovae are
    expected to have higher metallicity.  These higher metallicities are then
    preferentially passed on to new stars in those galaxies.  Old stars,
    necessarily produced in old galaxies, have low metallicity.  The 
    average atomic number in a white dwarf star should be an increasing
    function of the metallicity of its environment at the time of birth.
    
    In the susy phase transition model each white dwarf has a mass
    dependent lifetime as given in eq.\,\ref{PTprob}.  In a star-burst
    galaxy,  the supernova rate will preferentially come from stars with
    masses near the minimum lifetime.  Only when the star formation rate
    slows will stars with other masses contribute proportionately to the
    supernova rate.

    The above mentioned galactic properties have well established
    correlations with the SN Ia rate and the SN Ia peak brightness or
    Ni$^{56}$ production.

    \begin{table}[ht] \begin{center} \begin{tabular}{|c|c|c|c|c|}\hline
    Z&\,&\, Ni$^{56}$\;&Si$^{28}$\;&mean\\\hline 6&\,carbon fusion
    &\,0.00104\;&\;0.00081\;&\;0.00099\\\hline 8&\,oxygen fusion
    &\,0.00072\;&\;0.00051\;&\;0.00068\\\hline 10&\,neon fusion
    &\,0.00064\;&\;0.00042\;&\;0.00059\\\hline 12&\,magnesium fusion
    &\,0.00041\;&\;0.00020\;&\;0.00037\\\hline 
    \end{tabular}
    \end{center} 
    \caption{fraction of rest energy released in fusion
    reactions to Ni$^{56}$ and intermediate mass elements (IME)
    represented by Si$^{28}$ .  The last column gives the mean energy
    release per unit mass assuming equal production of Ni$^{56}$ and IME. }
    Each nickel nucleus decays to iron with a characteristic energy release
    which we neglect so the remaining energy release corresponds to that
    after standardization.  
    \label{Efusion} 
    \end{table}

    The best linear fit to the energy released in fusion to masses 
    $M(Ni^{56})$ and $M(Si^{28})$ is
    \be
    E_{fus} = (15.86 - 0.98 Z) M(Ni^{56}) + (13.43 - 0.95 Z) M(Si^{28})
    \label{Efus}
    \ee
    
    The observational results of ref.\,\cite{Konishi}, noted also in
    \cite{Calder},
     are that a mass of
    $Ni^{56}$ of about $0.55 M_\odot$ is produced in normal SN Ia roughly
    independent of metallicity though with large errors.  
    This estimate is also consistent with 
    the observations of \cite{Churazov} which, however, show some
    positive correlation of the nickel mass with the ejected mass. 
    Ref.\,\cite{Childress} supports a Ni$^{56}$
    production of about half of the ejected mass, $M$.
    For definiteness, we adopt, following \cite{Childress},
    \be
    M(Ni^{56}) &=& 0.5 M \quad .
    \ee
    We also assume slightly less than this in  
    intermediate mass elements, represented by $Si^{28}$ 
    \be
    M(Si^{28}) = 0.45\,M \quad .
    \ee
    This leaves an unburned mass
    \be
        M_{\mathrm unburned} \approx 0.05 \,M \quad .
    \label{unburned}    
    \ee
    The theoretical result
    of ref.\,\cite{breakdown} is that $M$ is bounded below by about
    $0.95\,M_\odot$.  
         
    The fact that the fusion energy available is maximized in carbon has
    suggested that, in the binary models, 
    white dwarfs with a higher percentage of carbon, as would
    be obtained in low metallicity environments, would have an increased
    production of nickel and, therefore, a greater luminosity \cite{Hoeflich} 
    but there are counter-arguments to this \cite{RopkeHillebrandt}. 

        It is commonly thought that the progenitors of normal type Ia
    supernovae are carbon-oxygen mixtures not because of any direct
    evidence \cite{Hillebrandt05} but because such mixtures lead to more
    successful standard model simulations once fusion is somehow
    triggered. As seen for example in \cite{SixIndications},  a roughly
    fifty-fifty mixture of carbon/oxygen fusing to a roughly fifty-fifty
    mixture of iron group elements and intermediate mass elements  will
    produce enough fusion energy to match experiment. In this case the
    maximum amount of energy input from a source beyond the standard
    model is sharply limited although some such energy input is needed to
    trigger fusion.   Typical white dwarfs will have maximum temperatures
    and pressures several orders of magnitude below that at which nuclei
    will touch and ignite fusion. Although, in \cite{breakdown},  no
    assumption was made as to the atomic composition of the white dwarfs
    undergoing supernova,  the progenitor (and ejected) mass was found to
    be in the range of O-Ne-Mg white dwarfs.

    In fact, an energy deficit in the standard model begins to open up if
    there are unburned remnants or if the initial carbon composition falls
    below about $0.5$.  Such low carbon content of at least some white dwarf
    stars is suggested by recent asteroseimological analyses
    \cite{Giammichele}. Since, contrary to expectations from simulations,
    a large central oxygen content 
    was found even for a light ($0.5\,M_\odot$) white dwarf, it would not be 
    surprising if, in heavy white dwarfs, a large O-Ne-Mg concentration
    becomes also observationally established. Furthermore, unburned oxygen 
    is observed in SN Ia but, normally, no unburned
    carbon.  This might be the first reason to doubt the progenitor
    identification as a C-O mixture.  The lighter elements in a mixture, 
    preferentially found in the low density outer regions of the star, are
    the most likely unburned ejecta.  Any significant amount of unburned
    elements, of course, reduces the fusion energy output and causes
    additional strain on the standard model.  
    Further supporting evidence comes from ref.\,\cite{Isern} 
    which finds that among
    carbon-oxygen white dwarfs the oxygen content near the center is nearly
    $80\%$ with the carbon content increasing toward the outer strata and
    these white dwarfs come from main sequence stars in the 5 to 7 solar
    mass range.  
    
    In the phase transition model \cite{breakdown} the progenitor mass
    ranges roughly from roughly $M_\odot$ to $1.4\,M_\odot$ which, 
    neglecting a small ($\approx 0.1\%$)
    surviving compact remnant, roughtly agrees with observations on the range
    of ejected mass \cite{Scalzo}.  White dwarfs in this mass range are known
    to be primarily oxygen-neon-magnesium mixtures or, at least, to have an
    oxygen-neon core \cite{Madej04}. Although not conclusive,  the above
    considerations support the hypothesis that the progenitors are primarily
    O-Ne-Mg.  This possibility has been studied in ref.\,\cite{Marquardt} 
    assuming some unknown external trigger.  However, in order to have 
    sufficient energy release from fusion alone, these models boost the 
    fraction of Ni$^{56}$ in the final state which then produces overly
    bright explosions.
    Very close to the Chandrasekhar mass there is a debated suggestion
    as noted above that the white dwarfs have an iron core which, 
    of course, produces no energy output from fusion but
    would produce a large amount of energy if there is degeneracy breakdown.

    Table\,\ref{Efusion} in connection with the energy sufficiency of
    carbon-oxygen dwarfs indicates that, if the progenitor nucleus has atomic
    number 8 or higher, the energy release from fusion alone is inadequate to
    explain observations and appreciable energy deposition is required from a
    source beyond the standard model.  This could lead to a problem
    in the binary models with understanding the homogeneity of SN Ia namely
    the total fusion energy available is sensitive to the composition of the
    white dwarf; dwarfs with a higher metallicity would produce a
    significantly smaller amount of Ni$^{56}$ and/or a significantly lower
    amount of ejecta kinetic energy contrary to observations.  On the other
    hand, if SN Ia come from a very limited range of white dwarf masses near
    $M_C$ , there is also a problem in producing a sufficient rate of
    supernovae. 

    As we show below, the phase transition model provides an extra energy
    input increasing with atomic number.   For example, the energies released
    if the protons in carbon or oxygen convert to scalars dropping into the
    susy ground state are 
    \be \Delta E(C) &=& 4 (E_{1p} - E_{1s}) + 6 (E_{1s}
    - \tilde{E}_{1s}) \\ 
    \Delta E(O) &=& 6 (E_{1p} - E_{1s}) + 8 (E_{1s} -
    \tilde{E}_{1s}) 
    \label{COenergies} 
    \ee 
    where $\tilde{E}_{1s}$ is the
    energy of the susy $1s$ state which may or may not be the same as that in
    normal carbon or oxygen.

    The nuclear shell model energies exhibiting the magic numbers have been 
    known since the 1950's, \cite{ShellModel} \cite{Bakken} and can be 
    used to determine the excitation 
    energies of various nuclei.  A modified harmonic oscillator potential
    or a modified square well illustrates the approximate equally spaced
    shells. 
    The harmonic oscillator parameters of the low lying elements, $6 \le Z
    \le 12$, depend on the atomic number but preserve the equal spacing so the
    excitation energies are scaled relative to those of carbon. 
    We can calibrate the excitation energies using the empirical
    relation between the $1s$ ground state to the first excited state in
    carbon and oxygen  \cite{tunl}: 
    \be\nonumber E_{1d5/2}(C) - E_{1s}(C) = 4.44 \,\mathrm{MeV}\\ 
    E_{2s}(O) - E_{1s}(O) = 6.05 \,\mathrm{MeV} \quad . 
    \ee
    The $2s$ state and the $1d5/2$ are nearly degenerate being part of the
    same shell model energy level. The total internal energies of carbon and
    oxygen relative to the $1s$ ground state, multiplying by a factor of 2 to
    include neutrons, are then as an example
    \be\nonumber 
    E_C &=& 2\cdot 4\cdot E_{1p3/2}(C) = 12.008\,\mathrm{MeV}\\ 
    E_O &=& 2\cdot (4\cdot E_{1p3/2}(C) + 2\cdot
    E_{1p1/2}(C)) \cdot 6.05/4.44 = 26.5\, \mathrm{MeV}\quad . 
    \ee
    In this article, given the present state of observations, we are
    interested primarily in the sense of the metallicity effects and only
    roughly in the magnitude of the effects. With this in mind we will assume
    \be E_{n,l}(Z+1) - E_{1s}(Z+1) \approx (E_{n,l}(Z) - E_{1s}(Z))\cdot
    6.05/4.44 \quad . \ee We can then construct table\,\ref{ExEnergies} for
    the energies of the low lying elements relative to the 1s ground states.
    \begin{table}[ht] 
    \begin{center} 
    \begin{tabular}{|c|c|c|}\hline
    Z&\,element\,&$\Delta \rho/\rho$\;\\\hline 6&\,carbon\,&$0.0467 =
    0.0078\cdot Z$ \\\hline 8&\,oxygen\,&$0.0562 = 0.0070 \cdot Z $\\\hline
    10&\,neon\,&$0.0719 = 0.0072 \cdot Z$\\\hline 12&\,magnesium\,&$0.0824 =
    0.0069 \cdot Z$\\\hline 26&\,iron\,&$0.204 = 0.0079 \cdot Z$\\\hline
    \end{tabular} 
    \end{center} 
    \caption{The fractional change in mass or
    density of the indicated element on giving up its excitation energy due
    to degeneracy breakdown as in the susy phase transition. The values for
    $\Delta\rho/\rho$ include an equal contribution from neutrons.  If there
    is an extra contribution coming from the second terms in
    eq.\,\ref{COenergies},  this would also be proportional to the atomic
    number.  We neglect the possible suggestion of quadratic $Z$ dependence.}
    \label{ExEnergies} 
    \end{table}

    The main point of table\,\ref{ExEnergies} (and of this article) 
    is that the susy energy
    release increases with atomic number while the energy from fusion shown
    in table\,\ref{Efusion} decreases with Z.  This complementarity is the
    beginning of the susy explanation for the uniformity of the
    supernovae Ia.

    In earlier work on the phase transition model based on the Fermi gas
    model for even-even nuclei,  the Pauli energy was taken to be simply and
    universally proportional to the density, $\Delta \rho \approx 0.02\rho$
    and a term proportional to a power of $\rho/\rho_c$ was added to the
    action to allow for the possibility of a transition-inhibiting effect
    analogous to the inhibiting of the liquid to gaseous transition at high
    pressure.   In the absence of such a term the action would go to zero as
    the white dwarf mass approaches $M_C$ in which limit the Coleman-DeLuccia
    model provides no guidance.   The white dwarf lifetime defined
    by eq.\,\ref{PTprob} would approach infinity as one
    approaches the Chandrasekhar mass from below.  However, the actual 
    lifetime above a mass of about $1.38\,M_\odot$ would approach zero due
    to standard model fusion ignition.

    In the nuclear shell model as tabulated in table\,\ref{ExEnergies},
    the Pauli excitation energies, neglecting a possible quadratic term,  are
    proportional to the atomic number: \be \Delta \rho \approx 0.007\,Z
    \,\rho \quad . \ee Thus in the Shell Model the critical density
    in eq.\,\ref{action} would be
    expected to be inversely proportional to the atomic number.   We would
    define a new critical density, $\rho_c$, and minimum lifetime, $\tau_0$,
    as free parameters in the fit to the delay time distribution.   The
    action is then parameterized as
    \be 
    B = (\frac{ \rho_c}{\overline{Z}\rho})^3 + b_0 (
    \frac{\overline{Z}\,\rho}{\rho_c})^{4/3} \quad . 
    \label{Zdep-action} 
    \ee 
    Since white dwarf masses in the range above $0.9$ are expected to be
    progressively O-Ne-Mg mixtures,
    we write for mean atomic number:
    \be 
    \overline{Z} = 7 + 10 \,(M - 0.9) \quad . 
    \label{Zbar} 
    \ee 
    The calculations, for simplicity,
    treat a star of fixed atomic number constituents assuming that
    interpolating between integer $Z$ is a reasonable approximation to a
    mixture of constituents.  The parametrization of eq.\,\ref{Zbar}
    ranges from an average atomic number between that of carbon and oxygen
    at $M=0.9$ to that of magnesium as $M$ approaches the Chandrasekhar mass.
    Some $25\%$ to $30\%$ of white dwarfs show metal lines in their spectra
    although how this composition is produced is debated.
    Perhaps, in an environment enriched
    by a history of nearby supernova, a main sequence star concentrates 
    higher $Z$ elements near the center
    and preferentially passes them on to the daughter white dwarf 
    after the lighter elements are blown off.
    It has also been suggested that spectral evidence for high Z elements 
    in cool white dwarfs could indicate accretion from rocky planets
    \cite{Kawka},\cite{Zuckerman}.

    With the Z dependent action of eq.\,\ref{Zdep-action}
    the inverse lifetime for a white dwarf of mass M is
    \be 
    \tau(M)^{-1} = \frac{1}{\tau_0} \frac{V(M)}{V_\mathrm{max}}
    \label{WDlifetime} 
    \ee 
     with now
     \be 
     V(M) = \int\,d^3r \, e^{-(\rho_c/(\overline{Z} \rho(r))^3-b_0\,(\overline{Z} 
     \rho(r)/\rho_c)^{4/3}} \quad . 
     \ee
    $V(M)$ is an increasing function of $\overline{Z}$ which would make 
    the lifetime
    a decreasing function of $\overline{Z}$ but this dependence is largely
    cancelled by the normalization to $V_{max}$.
    Relying on the observed lack of a compact remnant and observational 
    evidence ref.\,\cite{Gilfanov-Bogdan} that accretion is
    statistically not a major factor in white dwarfs undergoing SN Ia, we can
    neglect accretion and identify the ejected mass with the progenitor mass
    at birth. Multiplying by the white dwarf birth mass distribution from
    section 2 we can write the combined distribution in delay time and
    progenitor mass.
    \be \frac{d^2 N}{dt dM} = a_0 F(M) e^{-t/\tau(M)}\,/\tau(M) \quad .
    \label{d^2N} 
    \ee
    From eq.\,\ref{d^2N} we can derive the SN Ia rates as functions
    of delay time, $t$, since birth and ejected mass or birth mass. 
    These are 
    \be 
    \frac{dN}{dt} = a_0 \int dM\,F(M)\, e^{-t/\tau(M)}/\tau(M) 
    \label{DTD} 
    \ee 
    and 
    \be
    \frac{dN}{dM} = a_0 F(M)\cdot { (1 - e^{-t_g/\tau(M)})}\quad . 
    \ee
    It takes about $0.04 Gyr$ for a heavy main sequence star to produce
    a white dwarf so the time since parent star birth and the time since
    white dwarf birth (defined as the cessation of fusion) differ by
    about $0.04 Gyr$.  The $t$ in eq.\,\ref{DTD} should be replaced
    by $t-0.04$ if we want $t$ to represent the time since main sequence 
    star formation.
    
    \section[7]{The Delay Time and Ejected Mass Distributions}

\begin{figure} [!ht] \centering
    \includegraphics[scale=0.85]{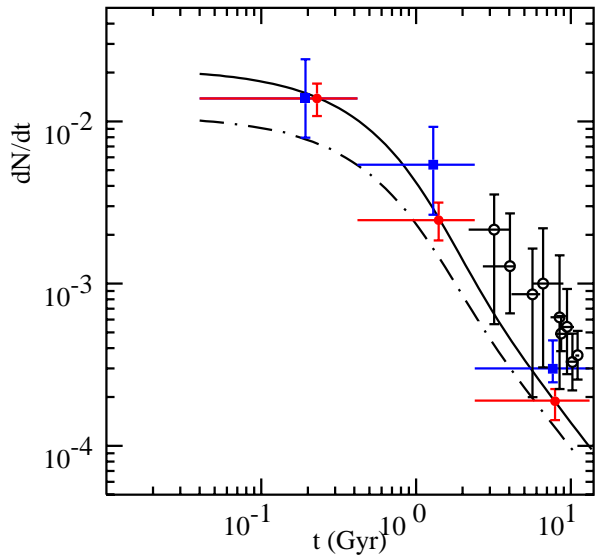} \caption{SN Ia delay time
    distributions.  The solid curve shows the DTD for average to high
    metallicity environments.   For low metallicity environments the DTD is
    reduced as shown in the dot-dashed curve due to the modified parameters
    of eq.\,\ref{Newb_1}.  Data is taken from \cite{MaozMannucciBrandt}
    neglecting earlier data with larger errors.} 
    \label{dNdt} 
    \end{figure}

       The distribution in delay times between white dwarf birth and SN
   Ia explosion has been measured with several techniques, the most
   accurate of which is illustrated in the red points of figure 1 of
   \cite{MaozMannucciBrandt}.  We do a $\chi^2$ minimization of
   eq.\,\ref{DTD} to these three large bin data points leading to the fit
   shown in figure\,\ref{dNdt}

    The resulting best fit parameters are 
    \be\nonumber 
    \rho_c &=& 58.8 \,M_\odot/{R_E}^3 = 4.50\,10^8\,g/\mathrm{cm^3}\\ \nonumber 
    \tau_0 &=& 0.418 \,Gyr\\ 
    b_0 &=& 0 \quad . 
    \ee

\begin{figure} [!ht] \centering
    \includegraphics[scale=0.85]{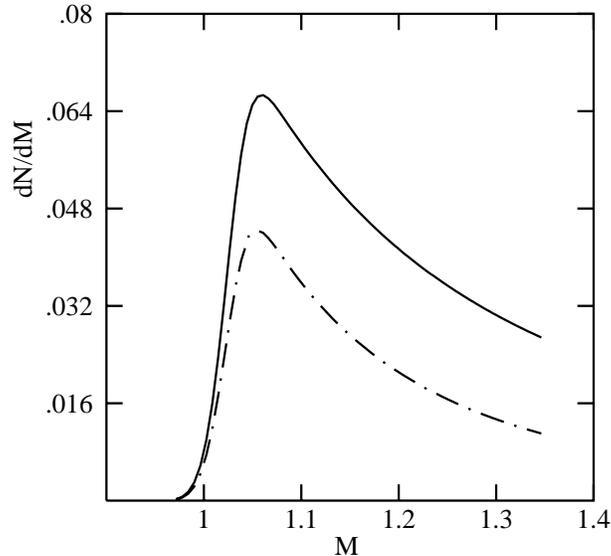}
    \caption{Progenitor mass distribution in high to average metallicity
    environments (solid curve) and in low metallicity environments
    (dot-dashed curve). Neglecting the compact surviving remnant these curves
    represent also the ejected mass distribution predictions.} \label{EJMD}
    \end{figure}

    Some discussion is warranted here.  First of all the best fit is not
    sharply determined.  Good fits with $\chi^2$ values less than 1 per
    degree of freedom are found with $\rho_c$ varying by up to $30\%$ and
    $b_0$ values as large as $1.5$. Correlated values of $\tau_0$ are found
    varying by as much as ten percent from the best fit.  With respect to the
    $b_0$ values one should note that the Coleman-DeLuccia theory, while not
    specifying non-leading behavior in the action,  is derived
    for large action.  With $b_0 = 0$, the action approaches zero near the
    center of high mass white dwarfs.    With $b_0 \approx 0.5$ or larger, the
    action is everywhere greater than unity.  However,  since the volume
    integral for $V(M)$ suppresses the behavior near $r=0$,  the theory should
    be acceptable even for $b_0 = 0$ especially since the predicted supernova
    progenitor mass distribution shown in figure\,\ref{EJMD} is suppressed at
    high mass.  In addition, as the white dwarf mass rises above $1.38$,
    nuclei will be within range of the strong interactions so fusion will
    be ignited preempting the susy phase transition and making the action $B$
    irrelevant.
    
    The white dwarf lifetime as a function of its mass is not greatly
    changed from the plot shown in ref.\,\cite{breakdown} where there is no
    explicit Z dependence in the action but, in the present work, the 
    free parameters have been adjusted accordingly. 

    The delay time distributions, DTD, have a shape that are nearly
    independent of the host galaxy metallicity although the integrated
    supernova rate is significantly lower in low metallicity environments.
    Moreover, supernovae in star-burst galaxies would be expected to have an
    enhanced rate especially at small delay times.  The shape and 
    magnitude of the DTD
    is well reproduced by the phase transition model which also gives a
    prediction for the behavior at small delay times.  The fact that, over a
    restricted range, the observed DTD plotted on a log-log scale is
    approximately linear with a slope near the dimensional analysis value
    of $-1$ has been widely cited as
    supporting the double degenerate model \cite{Maoz-Mannucci-Nelemans}.
    However, the binary models as reviewed there underpredict the DTD
    and vary among themselves by a factor of 3 to 10 in the low delay time
    region (see especially fig. 8 of reference\,\cite{Maoz-Mannucci-Nelemans}).

    In the phase transition model the ejected mass should be the same as
    the progenitor mass since the surviving susy core has a negligible mass.
    One sees from fig.\,\ref{EJMD} that the progenitor mass is strongly peaked
    near one solar mass.

    In the double degenerate model for SN Ia it has been suggested that
    the presence of a third star might be needed to throw two white dwarfs
    into each other with a short delay time.  This possibility reduces the
    predicted rate and has now been strongly disfavored
    as a major source of SN Ia \cite{Toonen}.

\begin{figure} [!ht] \centering
    \includegraphics[scale=0.85]{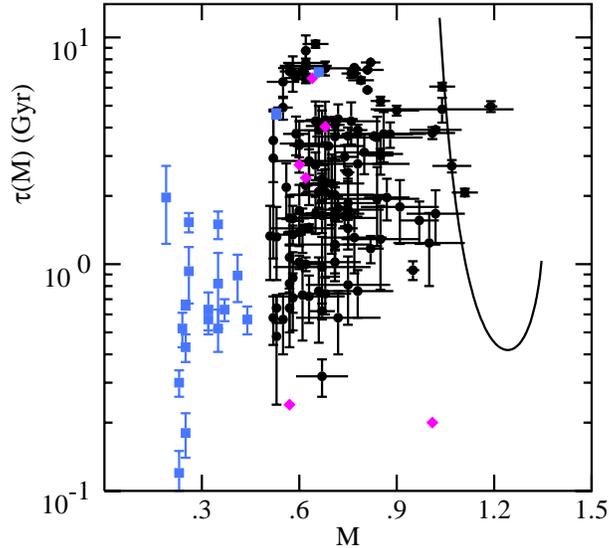}
    \caption{White Dwarf lifetime vs stellar mass in the susy phase
    transition model.  Of the $95$ cool DA white dwarfs in
    the data of \cite{Bergeron} those with mass greater than $0.5\,M_\odot$ 
    are shown in black with measured ages.  Of these $95$, $19\%$ ``known or 
    suspected" of being in double degenerate binaries are plotted in blue
    squares as is the one DB white dwarf in this sample.  The DB dwarfs
    have, by definition, a thin helium atmosphere. 
    The uncertainty in the mass of these stars which ranges from 
    $2\%$ to $50\%$ is ignored.  The present absence of high mass white dwarfs 
    in DD binaries is the major problem facing the DD scenario. 
    It is unknown why the DB stars are so much less likely to be in double
    degenerate binaries which could be a problem for the double detonation
    model in which fusion is ignited in an accreting helium layer.   
    The six nearest white dwarfs, shown in magenta,
    all have lifetimes greater than the current age of the universe.}  
    \label{Lifetime} 
    \end{figure}

    The white dwarf lifetime as a function of its mass is shown in
    fig.\,\ref{Lifetime}.  the phase transition model predicts no old white
    dwarfs in the quasi parabolic region except for exceptional, rapidly
    accreting stars as discussed in ref.\,\cite{breakdown}.  It should be
    noted that the best fit parameters from the delay time distribution also
    correctly predict the edge of the white dwarf age vs mass plot in 
    fig.\,\ref{Lifetime} as well as, roughly,
    the lower edge of the ejected mass distribution as measured in
    \cite{Scalzo}.

    In terms of the gravitational binding energy, $B$, (not to be 
    confused with the action of eq.\,\ref{action}), the fusion energy
    release, $E_{fus}$,  the electron internal energy released when the star
    explodes, $U$, and the kinetic energy of the ejecta, $E_{kin}$, we can
    write the total standard model energy deficit 
    \be 
     E_x = B + E_{kin} - E_{fus} - U  \quad . 
    \ee   
    These separate energies are calculated in
    ref.\,\cite{SixIndications} and elsewhere. 
    For instance, as treated there,
    \be
     E_{\mathrm{kin}} = 5.6\, 10^{-4} \,
     (1.22 M_\odot + 0.2 M) \quad .
    \label{Ekin}
    \ee
    In addition to the decay energy from nickel and other radioactive elements,
    a small amount of radiated energy, about $1\% E_{\mathrm{kin}}$, 
    associated with the production of
    this kinetic energy should be considered as included in $E_{\mathrm{kin}}$. 
    As alluded to above, the
    standard model energy deficit is consistent with zero in the case of a
    roughly fifty-fifty mixture of carbon and oxygen. However, since no
    compact remnants are observed and the ejected mass distribution ranges
    from $0.9\,M_\odot$ to $1.4\,M_\odot$ in which mass range stars are
    expected to be oxygen-neon-magnesium mixtures, there is a substantial
    energy deficit in the standard model which we propose is compensated by
    the susy energy release:     
    \be 
      E_{susy} = E_x \quad . 
    \ee     
    The total energy released beyond that required to unbind the star is 
    \be 
    E_t = E_{susy} + E_{fus} + U - B \quad .
    \label{Et} 
    \ee
    \begin{figure} [th] \centering
    \includegraphics[scale=0.85]{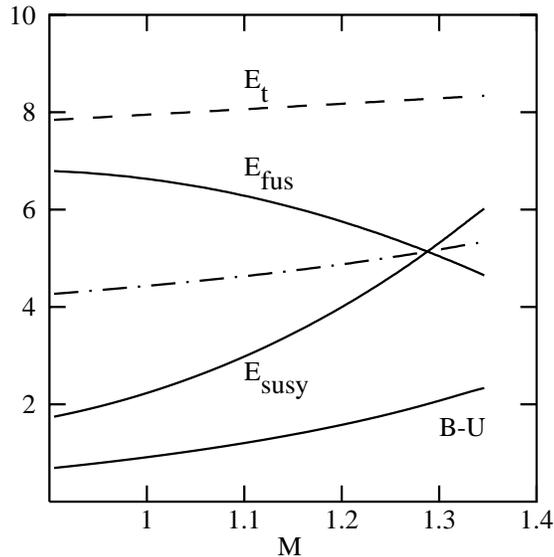}
    \caption{Various energies in SN Ia events in units of
    $10^{-4}\,M_\odot$ versus ejected mass, $M$, in solar mass units. The
    curves label the gravitational binding energy minus internal electron
    energy, $B-U$, the total fusion energy, $E_{fus}$, released and the
    nuclear degeneracy energy released in the small core, $E_{susy}$.  The
    mean of the fusion energy and released susy energy is shown in the
    dot-dashed curve. The total released energy, \ref{Et}, 
    beyond that required to unbind the star is shown in the dashed curve.} 
    \label{energies}
    \end{figure}
    As stated in the caption of table\,\ref{Efusion}, the energy
    release is that after standardization for the amount of nickel
    \cite{Phillips}
    and does not include the decay energy of Ni$^{56}$.
    Although the fusion
    energy release in the full star and the susy energy release in a small
    core vary rapidly with progenitor mass,  the sum of the two grows only
    slowly, closely tracking the total energy release beyond unbinding.  We
    propose that this is the basis for the uniformity of SN Ia events. The
    susy core which should shrink to a black hole after cooling has a mass
    roughly $0.1\%$  of the progenitor mass:
    \be 
    m_{core} = \frac{E_{susy}}{0.007\cdot \overline{Z}} \quad .
    \label{coremass} 
    \ee
    In fig.\,\ref{energies}, it is seen that the susy model predicts,
    after standardization,
    about a $5\%$ variation in total energy release beyond unbinding. 
   
    Since, after standardization, the energy release beyond
    unbinding rises toward higher white dwarf mass and mean atomic number,
    the model is consistent with the observation \cite{Gallagher}, \cite{Pan} 
    that the 
    Hubble residual correlates positively with metallicity.  At very low
    metallicity as would be obtained at high red shift, this effect would
    need to be taken into effect in determining the apparent Hubble constant.
    
    In the binary models the released energy tracks the more rapidly varying
    fusion energy making the Phillips relation and the SN Ia uniformity
    a long-standing challenge.        

    \section[8]{Summary}

    Some forty five years since the binary models for SN Ia were
    proposed, a review of all the binary models \cite{Livio}
    has still not been able to identify a
    unique progenitor system nor explain how a mixture of widely
    different progenitor systems can produce the observed homogeneity.  We
    have argued that the possibility of a phase transition mechanism should
    not be ignored.  Going beyond the six earlier indications of radical new
    physics in SN Ia \cite{SixIndications}, the host galaxy effects treated
    here could be taken as supporting the idea that the phase transition
    responsible for SN Ia is a  transition to an exact susy phase. This 
    possible connection is based on the indication that the standard 
    model energy deficit grows
    with progenitor mass and metallicity as does the degeneracy energy.

     With regard to the single degenerate model for SN Ia, continuing
     research on this model is based on the possibility of somehow
     avoiding the counter-indications from the X ray data of
     \cite{Gilfanov-Bogdan}. The $5\%$ Gilfanov-Bogdan limit is based on the
     inconsistency, given X ray constraints, of $1.2 M_\odot$ white dwarfs 
     accreting to the Chandrasekhar mass. The conflict is exacerbated if the
     progenitors are initially below one solar mass, as required if they
     are C-O dwarfs,  since then an even higher accretion rate would be
     required.

    In the double degenerate model, continuing research requires the
    expectation that a substantial population of high mass white dwarf
    binaries will be discovered above that suggested by the Salpeter initial
    mass function and significantly above the $20\%$ ``known or suspected" 
    in the sample of ref.\,\cite{Bergeron}. 

     In the models of Sim et al. \cite{Sim} the trigger is
   assumed to provide negligible energy release to supplement the
   available fusion energy.  We have argued that the actual composition
   should be O-Ne-Mg with the reduction in available fusion energy being
   compensated by the degeneracy energy released in a small
   core.  The peak in the progenitor mass distribution, shown in 
   figure$\,\ref{EJMD}$, is
   in reasonable agreement with one of the Sim et al. models and the
   range of progenitor masses roughly agrees with observations
   \cite{Scalzo}. The delay time distribution is also well reproduced as
   in fig.\,\ref{dNdt}.  Although we have found in ref.\,\cite{breakdown}
   that accretion, in the susy model, is generally not important in 
   supernova production,
   a small accretion rate ($\approx 10^{-11} M_\odot /yr $) 
   onto high mass white dwarfs could be effective
    in producing the $1\%$ of super-Chandrasekhar events 
    and the second peak in the ejected mass distribution possibly suggested 
    in ref.\,\cite{Scalzo}.  The super-Chandrasekhar events, of course,
    would require some temporary stabilization effect such as the 
    spin-up/spin-down model \cite{DiStefano}.  
    The tail of the Salpeter prediction of the 
    white dwarf mass distribution above a mass of $1.38$ will also 
    immediately produce a supernova since, above this mass, the nuclei will 
    be within the range of the strong interactions.  This could also 
    contribute up to $1\%$ of supernovae.
   
    In the ejected mass distribution and the delay time distribution,
    the metallicity dependence should be visible in the future high
    statistics supernova data expected from the Gaia collaboration which
    should also settle the question as to the possible existence of high
    mass double degenerate systems \cite{Brevik}.   Gaia has already
    succeeded in identifying binary companions among the progenitors of
    core collapse supernovae \cite{Boubert} but has not yet reported
    observations of binary progenitors of SN Ia.  When the double mass
    distribution and other orbital distributions of white dwarf binaries
    are in hand, the DD scenario, if correct, should be able to produce 
    delay time distributions and ejected mass distributions to compare 
    with those of the phase transition model. 

    The minimum energy released in the nucleation of a critically
    sized bubble is 
    \be 
    E_{susy}(min) = {\overline{\Delta \rho}} \cdot
    \frac{4 \pi}{3} {R_c}^3 = 36 \pi^3 S^3/{\overline{ \Delta \rho}}^2
    \label{Emin} 
    \ee     
    where $R_c$ is the critical radius of eq.\,\ref{CritRad}. The fit to 
    the critical density implies that the surface tension is    
    \be 
    S \approx 2\cdot 10^{-21} M_\odot/{R_E}^2 \quad . 
    \ee 
    The minimum bubble size from
    eq.\,\ref{CritRad} is then small on the stellar scale but 
    large compared to the inter-atomic scale.  
    The much larger full susy energy released
    depends on the final radius of the susy bubble.
    
    The growth of the susy bubble is halted when the critical radius
    becomes greater than its current radius as could happen due to a sharp
    falloff of density or to its entrance into a region of lower atomic
    number.  The supernova explosion is expected to create a cavity of
    density significantly lower than that of interstellar space. The extent
    of a white dwarf iron core as discussed above, could also be critical 
    since iron is inert to
    fusion but would provide a large energy release if there is degeneracy
    breakdown as seen in table\,\ref{ExEnergies}.  In eq.\,\ref{Zbar} 
    we have tentatively assumed that
    the mean atomic number never increases above that of magnesium even if
    a small iron core builds up in the center.  
    
    If, on the other
    hand, the iron core becomes a non-negligible fraction of the total,
    fusion will be greatly suppressed and the 
    observed sub-luminous events (SN Iax) may be produced.  These would be
    predicted to have ejected mass near the Chandrasekhar limit and might
    be contained in the high ejected mass events of ref.\,\cite{Scalzo}. 

    Prior to consideration of the host galaxy correlations, indications
    of the need for radical new physics could be satisfied by other types of
    phase transitions such as transitions to a quark-gluon plasma although no
    such model has as yet been worked out for SN Ia. The observation that the
    explosion energy increases with metallicity as does the degeneracy
    energy is in line with exact susy being the phase transition final state.
    Finally, it must be noted that the hypothesis of a phase transition to an
    exact susy phase implies that the broken susy phase must also exist and
    susy particles must be found at sufficiently high accelerator energies if
    the model suggested here is to be viable.

\end{document}

%% file: defs.tex
\usepackage{graphics}
\usepackage[dvips]{color}
\usepackage{multicol}

\usepackage{amsmath,color,fancybox,graphics,graphpap,rotating}
\definecolor{white}{rgb}{1,1,1}
\definecolor{yellow}{rgb}{0.95,0.75,0.1}
\definecolor{red}{rgb}{0.5,0,0}
\definecolor{green}{rgb}{0,1,0}
\definecolor{blue}{rgb}{0,0.5,1}
\definecolor{bgcolor}{rgb}{0.94,0.91,0.78}

\definecolor{lblue}{rgb}{0,0.8,1}
\definecolor{dblue}{rgb}{0,0,.6}
\definecolor{dgreen}{rgb}{0,0.3,0}
\definecolor{lila}{rgb}{0.8,0,0.8}
\definecolor{violet}{rgb}{1,0,1}
\definecolor{grey}{rgb}{0.3,0.3,0.3}
\definecolor{turquoise}{rgb}{0,.9608,1}

\def\lsim{\raise0.3ex\hbox{$\;<$\kern-0.75em\raise-1.1ex\hbox{$\sim\;$}}}
\def\gsim{\raise0.3ex\hbox{$\;>$\kern-0.75em\raise-1.1ex\hbox{$\sim\;$}}}

\newcommand{\nee}{\nonumber \end{eqnarray}} 
 
\usepackage[dvips]{color}
\usepackage{multicol}
\usepackage{amsmath,color,fancybox,graphics,graphpap,rotating}
\def\lsim{\raise0.3ex\hbox{$\;<$\kern-0.75em\raise-1.1ex\hbox{$\sim\;$}}}
\def\gsim{\raise0.3ex\hbox{$\;>$\kern-0.75em\raise-1.1ex\hbox{$\sim\;$}}}

\usepackage{amsmath,color,fancybox,graphics,graphpap,rotating}
\newcommand{\be}{\begin{eqnarray}}
\newcommand{\ben}{\begin{eqnarray}\nonumber}
\newcommand{\ee}{\end{eqnarray}}

\definecolor{lred}{rgb}{1,0.3,0.3}
\definecolor{red}{rgb}{0.5,0,0}
\definecolor{dblue}{rgb}{0,0,.6}

\def\ApJ{{ Astrophys. J.} }
\def\ApJL{{ Astrophys. J. Letters} }

\def\AA{{ Astron. \& Astroph.} }

\def\JCAP{{\it Journ. of Cosmol. \& Astropart. Phys.}Ê}

\def\MNRAS{{ Month. Not. Roy. Astr. Soc.} }
\def\Nature{{ Nature} }

\def\PRD{{ Phys. Rev.} {\bf D} }

\def\RMP{{ Rev. Mod. Phys.} }